\documentstyle[preprint,aps]{revtex}
\begin{document}
\draft
\preprint{UTPT-97-06}
\title{Stochastic Gravity and Self-Organized Critical Cosmology}
\author{J. W. Moffat}
\address{Department of Physics, University of Toronto,
Toronto, Ontario M5S 1A7, Canada}

\date{\today}
\maketitle

\begin{abstract}%
A stochastic theory of gravity is described in which the metric tensor is a random variable
such that the spacetime manifold is a fluctuating physical system at a certain length scale.
A general formalism is described for calculating probability densities for gravitational
phenomena in a generalization of general relativity (GR), which reduces to classical GR
when the magnitude of the metric fluctuations is negligible. Singularities in gravitational collapse
and in big-bang cosmology have zero probability of occurring. A model of a self-organized
critical
universe is described which is independent of its initial conditions.
\end{abstract}
\vskip 0.3 true in
To be published in the Proceedings of the workshop: {\it Very High Energy Phenomena in the
Universe}
of the XXXIInd Rencontres de Moriond, Les Arcs, France, January, 1997.
\pacs{ }

{\bf  Introduction}

Classical general relativity is based on the assumption that the spacetime manifold is $C^2$
smooth down to zero length scales. This assumption is basically the source of the singularity
theorems of Hawking and Penrose\cite{Hawking}, which state that in GR a singularity must be
the final
state of gravitational collapse and that the big-bang must have occurred as a naked singularity in
cosmology, provided the strong and weak energy positivity conditions are satisfied. From our
knowledge of physical systems in Nature, through studies of physics, chemistry and biology 
we know that systems display noise at some length scale in the form of random fluctuations or
$1/f^a$ flicker noise or chaotic noise. If we consider spacetime geometry to be a physical system
that interacts with matter, then we should expect that it also will have metric fluctuations
corresponding to ``noise" in the spacetime geometry. Indeed, the assumption that {\it spacetime
geometry is smooth down to zero length scales seems quite unrealistic}.

Wheeler\cite{Wheeler} suggested some time ago that at the Planck length, $L_P=
\sqrt{{\hbar}G}=1.6\times
10^{-33}$ cm spacetime geometry would possess quantum fluctuations with metric fluctuations
of order, $\Delta g\sim 10^{-20}$. These are so small that we can ignore them for experimentally
accessible physical observations. Quantum gravity describes a strongly coupled phase in which
there are no correlations on length scales larger than $L_P$. However, we can consider
that spacetime fluctuations occur not involving Planck's constant $\hbar$, which display
correlations at much larger length scales and for self-organized critical gravitational phenomena
display correlations at all length scales. Theoretical and experimental studies have shown that
non-linear physical systems in ``noisy" environments show surprising behavior that does not
conform
to common intuitive experiences. Thus, we can expect that since GR is a highly non-linear theory
the fluctuating spacetime manifold when coupled to gravitational matter exhibits a much richer
behavior than is possible in a purely deterministic theory. A gravitational theory based on
statistical mechanics can display two kinds of systems: (1) a fine-tuning of parameters that leads
to critical points and phase transitions and (2), self-organization which can often describe critical
systems with structure spread out over every available scale. Such properties can be applied to
our understanding of the universe at both small and large scales. Indeed, it would be surprising if
non-linear gravitational physics is not a rich source of new phenomena determined by
non-linear complexity theory.

{\bf  Stochastic Gravity}

A probabilistic interpretation of spacetime has been developed based on defining the spacetime
metric tensor as a random stochastic variable\cite{Moffat1}. The probability that the metric $g$
takes
a value between $g$ and $dg$ is given by $P(g)dg$ where $P(g)$ is a probability distribution
defined as a scalar quantity with $P(g) \geq 0$ and normalized to unity on its range. A  canonical
formalism based on a $(3+1)$ foliation of spacetime in GR$^{(3)}$ leads to a Langevin equation
for the canonically conjugate momentum variable $\pi^{\mu\nu}$ treated as a stochastic variable.
The gravitational constant $G$ is defined to be a control parameter with the decomposition:
\begin{equation}
G_t=G+\sigma\xi_t,
\end{equation}
where $G$ is the average value of Newton's gravitational constant and $\xi_t$ is Gaussian
white noise with $<\xi_t>\equiv E\{\xi_t\}=0$ where $E$ denotes the expectation value;
$\sigma$ measures the intensity of the geometrical fluctuations of the metric. Thus, $G_t$ will
have a bell-shaped curve distribution, peaked at the average value, $G$. The stochastic
differential equation for the dynamical random variable $\pi_t^{\mu\nu}$ has the form:
\begin{equation}
\partial_t\pi_t^{\mu\nu}=f_t^{\mu\nu}+8\pi GT^{\mu\nu}+8\pi\sigma\xi_t T^{\mu\nu},
\end{equation}
where $f^{\mu\nu}_t$ is determined by terms involving the $(3+1)$ projected curvature tensor, 
$\pi_t^{\mu\nu}$ and the lapse and shift functions, $T^{\mu\nu}$ is the projected stress
energy-momentum tensor for matter
with the components $T_{\perp\,\perp}=T_{\mu\nu}n^\mu n^\nu$ and $T^\nu_{\perp}=
{h^\nu}_\alpha T^{\alpha\beta}n_\beta$; $h_{\mu\nu}=g_{\mu\nu}+n_\mu n_\nu$ is the
induced
spatial metric and $n^\mu$ is the normal unit vector to the Cauchy spacelike surface, $\Sigma$.
A Fokker-Planck equation can be derived for the probability density, $p(\pi_t)$:
\begin{equation}
\partial_tp(y,t\vert\pi_t,0)=-\partial_t[F_t(y)p(y,t\vert\pi_t,0)]+32\pi^2T^2\partial_{yy}
p(y,t\vert\pi_t,0),
\end{equation}
where $\pi$ and $T$ denote for convenience the canonically conjugate momentum
$\pi^{\mu\nu}$ and the
stress energy-momentum tensor $T^{\mu\nu}$, respectively.
This equation has an exact solution for a stationary probability density obtained for a
gravitational
system after a long time has elapsed:
\begin{equation}
p_S(\pi)={C\over 64\pi^2 T^2}\exp\biggl({1\over 32\pi^2\sigma^2T^2}
\int^\pi F(u)du\biggr),
\end{equation}
where $C$ is a normalization constant.

We can formulate a stochastic differential equation for the geodesic motion of a test particle:
\begin{equation}
du^\mu_s+\Gamma^\mu_{s,\alpha\beta}u^\alpha_s
u^\beta_s=-\zeta_s\Gamma^\mu_{s,\alpha\beta}u^\alpha_s u^\beta_s ds,
\end{equation}
where $\zeta_s$ is a Brownian motion process in terms of the proper time $s$, the Christoffel
symbol $\Gamma^\mu_{s,\alpha\beta}$ is treated as a random variable determined by the
stochastic metric $g_{s,\mu\nu}$, $u^\mu=dx^\mu/ds$ denotes the time-like four-velocity,
and $u_s^\mu$ describes this four-velocity as a random variable.
At the length scale for which the fluctuations of spacetime are significant, we can picture a test
particle moving in spacetime along a Brownian motion path such that $u_s^\mu$ does not have a
well defined derivative with respect to $s$ at a point on the world line. For large enough
macroscopic length scales for which the spacetime fluctuations can be neglected, the motion
of the test particle becomes the same as the deterministic geodesic equation of motion in GR.

A stochastic Raychaudhuri equation can be derived\cite{Moffat1} from which we can deduce
that caustic singularities in the spacetime manifold can be prevented from occurring if the
intensity
of fluctuations is big enough for a given length scale, assuming the standard positive energy
conditions. In the limit of classical GR, the Hawking-Penrose singularity theorems will continue
to
hold, for the Brownian motion fluctuations of spacetime are negligible and can be neglected.

Consider the case of inward radial motion in a Schwarzschild geometry. For radial motion we
have
\begin{equation}
{dt\over ds}=\biggl(1-{2GM\over r}\biggr)^{-1},\quad\quad {dr\over ds}
=-\sqrt{2GM\over r},
\end{equation}
where $M$ denotes the mass of the central particle. The stochastic differential equation for the
random variable $r_s$ is given by
\begin{equation}
dr_s=\sqrt{G}f(r_s)ds+\beta(\sqrt{G}+\zeta_s)f(r_s)ds,
\end{equation}
where 
\begin{equation}
f(r_s)=-\sqrt{2M\over r_s}
\end{equation}

and $\beta$ is a non-linear function of the Wiener process $\zeta_s$. By taking the
white-noise Gaussian
limit for short correlation times\cite{Moffat1}, we get the stationary probability density:
\begin{equation}
p_S(r)={Cr\over 2M}\exp\biggl(-{2\sqrt{2}\sqrt{G}\over 3\sqrt{M}\sigma^2}r^{3/2}\biggr),
\end{equation}
where $C$ is a normalization constant. We see that $r\rightarrow \infty$ is a natural boundary:
both the drift and the diffusion coefficients vanish as
$r\rightarrow\infty$. For
$r\rightarrow 0$ we have $p_S(r)\sim 0$. We therefore arrive at the result that as the particle
falls towards the origin there is zero probability for $r(s)$ to have the value zero, and consequently
there is zero probability of having a singularity at $r=0$ in the Schwarzschild solution. The
spacetime metric fluctuations smear out the singularity at $r=0$. 

An analysis of the collapse of a spherically symmetric dust cloud in our stochastic gravity theory
shows that a simililar result follows\cite{Moffat1}. As the star collapses there is zero probability
for the
star having a singularity as the final state of collapse. Moreover, it was also found that in the
gravitational collapse of a star there is zero probability of a black hole horizon forming with an
infinite red shift. The spacetime fluctuations become extremely intense as
$r\rightarrow 2GM$ and they quench the infinite red shifts. If we ignore the spacetime
fluctuations, then the standard GR results follow: both the singularity and the infinite red shift
black hole horizon must occur when a trapped surface forms during collapse. However, this
assumes that the spacetime geometry is perfectly smooth in the limit of zero distance scales and
infinite frequencies of red shifts- an assumption which seems contrary to all our experiences of
physical systems! 

An application of stochastic gravity to cosmology yields the result that the probability of a
singularity occurring at the big-bang is zero\cite{Moffat1}; the spacetime metric fluctuations
smear out
the singularity at $t=0$.

{\bf Self-organization and Criticality in Cosmology}

A major problem in modern cosmology is this: How could the universe evolve during more than
10 Gyr and become so close to spatial flatness and avoid the horizon problem? Why is the
universe so homogeneous and isotropic? How could such a critical state of the universe come
about without a severe fine tuning of the parameters?

The usual explanation for these questions is based on the idea of inflation\cite{Guth}. However,
inflation is a type of phenomenon that in statistical mechanics corresponds to the existence of an
attractor that requires {\it fine tuning of the parameters}. Indeed, in all models of inflation
introduced to date there exists fine tuning of the parameters of the models. This is the intrinsic
weakness of these models.

Recently, it has been proposed that the universe evolves as a self-organized critical system with
the expansion of the universe undergoing ``punctuated equilibria"\cite{Moffat2} with energy
being
dissipated at all length scales. The idea of a self-organized critical state was introduced by Bak,
Tang and Wiesenfeld\cite{Bak1} using a sandpile or a system of coupled damped
pendula as models. This remarkable idea has been applied to many different physical systems in
condensed matter physics, biology, earthquake studies and economics among others\cite{Bak2}.

We postulate that the universe evolves as a self-organized critical system in which 
the forces of expansion increase the cosmic scale $R$ in steps with the spacetime fluctuations
and
the increments of expansion $R_n$ sliding back in ``avalanches" which cannot be detected by the
local observer. The universe evolves towards a critical state with the individual spacetime
fluctuations and $R_n$ developing highly cooperative effects. The $R_n$ (or the density profile
$\Omega_n$) satisfies a nonlinear discretized diffusion equation with a threshold condition
associated with the 
critical value $R_c$ or $\Omega_c$. The dynamics leads to a stable state at some time $t=t_S$
{\it completely independent of how the universe began}. We can randomly add expansion,
$R_n\rightarrow R_n+1$, and induce slides $R_n\rightarrow R_n-1$ with a random distribution
of critical expansion differences and a uniformly increasing slope of expansion. The power
spectrum for the metric fluctuations $S(\Delta g)$ satisfies a power law
\begin{equation}
S(\Delta g)\sim (\Delta g)^{-\beta}
\end{equation}
associated with a self-similar fractal behavior. 

We are unable to predict the self-organized critical value for the density profile,
$\Omega=\Omega_c$ (observationally $\Omega_c\leq 1$), without further dynamical input from
GR. But we can explain with this model how the universe has to evolve to a stable critical value
$\Omega=\Omega_c$, which is independent of the initial conditons and without fine tuning of
the
parameters. The metric fluctuations display $1/f$ flicker noise, correlations of fluctuations occur
at all length scales,  and the universe evolves at the ``edge of chaos". There is only one possible
stable choice  (i.e., stable under local perturbations) for the present expanding universe whatever
its initial conditions. 

According to our assumptions the spacetime geometry fluctuates randomly at some length scale.
If we assume that the metric fluctuations are very intense at the beginning of the universe, and
that they smear out the light cones locally, then for a given short duration of time $\Delta t$ after
the big-bang there will be communication of information ``instantaneously" throughout the
universe. This will resolve the ``horizon" problem and explain the high degree of isotropy and
homogeneity of the present universe.

\acknowledgments

I thank the Natural Sciences and Engineering Research Council of Canada for the support of this
work.

\end{document}